\documentclass[conference]{IEEEtran}
\IEEEoverridecommandlockouts
\usepackage{cite}
\usepackage{amsmath,amssymb,amsfonts}
\usepackage{algorithmic}
\usepackage{graphicx}
\usepackage{textcomp}
\usepackage{xcolor}
\def\BibTeX{{\rm B\kern-.05em{\sc i\kern-.025em b}\kern-.08em
    T\kern-.1667em\lower.7ex\hbox{E}\kern-.125emX}}

\usepackage{mdframed}
\usepackage{xcolor}
\usepackage{soul}
\definecolor{darkgray}{rgb}{0.2, 0.2, 0.2}
\newcommand{\graycomment}[1]{%
    \vspace{0.5\baselineskip} 
    \textcolor{darkgray}{\textit{#1}}%
    \vspace{0.5\baselineskip} 
}

\begin{document}

\title{A Qualitative Investigation to Design Empathetic Agents as Conversation Partners for People with Autism Spectrum Disorder
\thanks{This research was funded in whole or in part by the Austrian Science Fund (FWF) [I 6465-B], under the frame of ERA PerMed.}
}

\author{\IEEEauthorblockN{1\textsuperscript{st} Christian Poglitsch}
\IEEEauthorblockA{\textit{ISDS - Institute of Interactive Systems and Data Science} \\
\textit{University of Technology Graz}\\
Graz, Austria \\
christian.poglitsch@tugraz.at}
\and
\IEEEauthorblockN{2\textsuperscript{rd} Johanna Pirker}
\IEEEauthorblockA{\textit{Media Informatics Group} \\
\textit{Ludwig-Maximilians University Munich}\\
Munich, Germany \\
johanna.pirker@ifi.lmu.de}
}

\IEEEoverridecommandlockouts
\IEEEpubid{\makebox[\columnwidth]{ 979-8-3503-5067-8/24/\$31.00~\copyright2024 IEEE \hfill} 
\hspace{\columnsep}\makebox[\columnwidth]{ }}
\IEEEpubidadjcol

\maketitle

\begin{abstract}
Autism Spectrum Disorder (ASD) can profoundly affect reciprocal social communication, resulting in substantial and challenging impairments. One aspect is that for people with ASD conversations in everyday life are challenging due to difficulties in understanding social cues, interpreting emotions, and maintaining social verbal exchanges. To address these challenges and enhance social skills, we propose the development of a learning game centered around social interaction and conversation, featuring Artificial Intelligence agents. Our initial step involves seven expert interviews to gain insight into the requirements for empathetic and conversational agents in the field of improving social skills for people with ASD in a gamified environment. We have identified two distinct use cases: (1) Conversation partners to discuss real-life issues and (2) Training partners to experience various scenarios to improve social skills. In the latter case, users will receive quests for interacting with the agent. Additionally, the agent can assign quests to the user, prompting specific conversations in real life and providing rewards for successful completion of quests.

\end{abstract}


\section{Introduction}

\begin{figure}[!t]
\centering
\includegraphics[width=8.5cm]{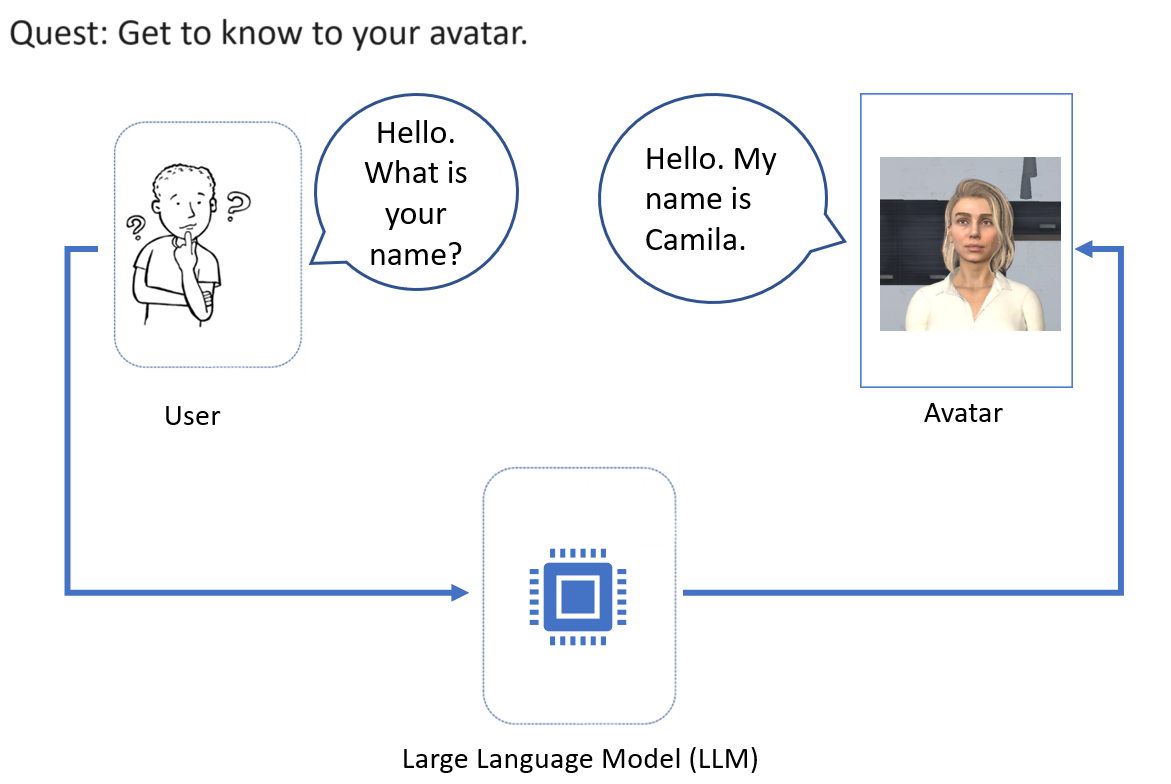}
\caption{\label{fig_overview} Proposed framework for initiating conversations with an agent, incorporating a starting quest.}
\end{figure}

Navigating through social interactions and everyday scenarios can present challenges for people with Autism Spectrum Disorder (ASD). According to the American Psychiatric Association \cite{AmericanPsychiatricAssociation}, core symptoms include: (1) impaired social perception (perspective-taking), (2) impaired interpersonal social competence (nonverbal communication, gaze behavior, gesture, facial expression), and (3) compulsive, stereotyped thinking and behavior, along with narrowly focused interests. ASD is becoming increasingly prevalent, and the challenge of 'social blindness,' characterized by difficulty recognizing emotions in others, poses significant barriers to people with autism in their daily lives.
With appropriate support and interventions, such as social skills training and the utilization of visual aids or assistive technologies, people with ASD can enhance their conversational abilities and navigate social interactions more effectively \cite{Karami2023}, \cite{MesaGresa2018}. Large Language Models (LLMs) offer continuous access to vast amounts of information and generate responses that simulate human-like interactions across a wide range of topics. This continuity is crucial to maintain learning success, as people with ASD often rely on memorizing typical scenarios to navigate social situations. In this paper, our objective is to design empathetic agents aimed at assisting people with ASD in honing their social skills and conversational abilities within a secure environment. The target audience of our tool is adults with ASD. The planned framework employs Unity for rendering and user interaction, utilizes avatars capable of displaying emotions from Reallusion, and integrates a LLM for chat completion, as illustrated in the figure \ref{fig_overview}. 

As a first step, we plan to conduct a qualitative investigation to gather requirements for social training for people with ASD.

\textit{Research questions}
\begin{itemize}
\item RQ1: What characteristics should empathetic agents have?
\item RQ2: How to visualize empathetic agents?
\item RQ3: How to interact with empathetic agents?
\item RQ4: How to evaluate success of empathetic agents?
\end{itemize}

\section{Related Work}

The potential of using virtual agents for use in assistive learning has been highlighted by various researchers. Systematic reviews demonstrate significant improvements in social skills, emotional abilities, daily skills, communication skills, and attention when interacting with agents, as evidenced in \cite{Karami2023} and \cite{MesaGresa2018}. The potential of using computer vision and graphics techniques to track a player's speech rate, facial features, eye contact, audio communication, and emotional states, as well as how personalization enhances learning motivation, is discussed in \cite{Ng2018}. Furthermore, games that are both enjoyable and aligned with evidence-based therapies reduce the gap between the recommended amount of therapy for children with ASD and the amount they actually receive \cite{Hiniker2013}. 

A model of empathy in virtual agents proposed by \cite{Rodrigues2014}, is based on psychological theories of empathy. The model treats empathy as a process modulated by the relationship between the agent and the user. In the work of \cite{Milne2010} an autonomous agent system can reproduce natural-looking facial expressions. An A/B evaluation revealed significant improvements in children who underwent this training. The study by \cite{Ali2020} highlights the importance of informing users about the purpose and limitations of the system, emphasizes that the realistic appearance and responsiveness of agents are essential for user engagement and underscores the importance of personalization. Agent requirements include serving as a tutor to improve and support understanding of language and conversation, maintaining a pedagogical focus, and being responsive \cite{Bernardini2021}.

LLM show the potential to simulate daily life of human beings. The evaluation results show that generative agents \cite{park2023generative} can produce individual and emergent social behavior with planning, scoring, and reflecting phases. These phases are vital modules for recalling past conversations and facilitating proactive agents that can initiate actions based on user interactions. This also enables the agent to proactivly assign quests to the user.
Further studies demonstrate that LLM are capable of collaborating effectively with humans. The proposed approach, called "ProAgent" \cite{Zhang2024}, underwent evaluation through experiments, demonstrating successful collaboration between the agent and humans.

The related literature indicates the potential for training social skills within a virtual environment and underscores the role of LLMs as proxy for an human conversation partner. Our paper contributes by addressing the gap in a qualitative investigation on utilizing generative agents for conversation training among people with ASD in a gamified scenario.

\section{Method}

Our contribution lies in designing empathy agents that aim to train social skills, with a particular focus on people with ASD. Our approach involves creating a questionnaire, identifying experts, conducting interviews with them, and subsequently analyzing the gathered data.

\subsection{Develop questionnaire}

As a first step, we developed a structured interview protocol comprising open-ended questions aimed at eliciting detailed responses from experts:

\begin{mdframed}
\begin{enumerate}
    \item Background and expertise
    \begin{enumerate}
        \item Do you have experience in the field of autism, and if so, what kind?
        \item What therapeutic methods do you know for promoting social skills in autism?
    \end{enumerate}    
    \item Design aspects
    \begin{enumerate}
        \item What characteristics should empathetic agents have to promote effective social interaction in people with ASD? Conversely, what characteristics should they not have?
        \item How should the personality of empathetic agents be designed to promote positive interaction?
        \item How should the background story of empathetic agents be designed to promote positive interaction?    
    \end{enumerate}    
    \item How should users and agents communicate with each other?    
    \item Visualization
    \begin{enumerate}
        \item Which human features should be emphasized in the visualization of empathetic avatars?
        \item Should these agents be realistically or stylized, and why?
    \end{enumerate}  
    \item Feedback and adaptation
    \begin{enumerate}
        \item What mechanisms can be introduced to give feedback to people with ASD on their interactions with empathetic agents?
        \item How can agents be adapted to individual progress and user needs?
    \end{enumerate}
    \item Evaluation
    \begin{enumerate}
        \item What metrics are important for evaluating the effectiveness and user satisfaction with social, empathetic agents?
        \item How should these agents be evaluated in user studies to measure their impact on learning social interactions?
    \end{enumerate}
\end{enumerate}
\end{mdframed}

\subsection{Recruit Experts}

We recruited experts in the autism field, including therapists, doctors, and clinical psychologists, who possess specialized knowledge and experience in working with people on the autism spectrum.  

\subsection{Data Analysis}

To evaluate the results of the interviews, we employed a qualitative analysis approach. This involved transcribing the recorded interviews verbatim and then systematically reviewing and coding the data to identify recurring themes, patterns, and insights. Through this process, our goal was to gain a comprehensive understanding of the perspectives, experiences, and recommendations shared by the interview partners. 

\section {Results}

In this section, we present the findings of our qualitative investigation. We successfully recruited seven experts for interviews, each lasting between thirty to fifty minutes. Throughout these interviews, we delved into various aspects of our research questions, gaining valuable insights from the diverse perspectives of the experts.

\subsection{Background of recruited experts}

Experts were recruited from medical centers, autism centers, and through contacts with clinical psychologists. We successfully recruited seven experts as interview partners, including two people with ASD, who are team members of partner groups under the ERA PerMed framework. The other experts are professionals in the fields of autism or clinical psychology. Three interviews were conducted in person, while the rest were conducted online. The ages of the experts range from twenty-two to fifty-five, including one male and six female experts. All experts provided their informed consent for the interview. 

\subsection{Design Aspects}

This section delves into the design aspects of agents, taking into account their characteristics, personality traits, and background. According to the experts, two potential solutions have emerged: (1) an agent functioning as a supportive friend, offering constructive feedback and guidance, and (2) a training partner specifically for simulating everyday scenarios. 

\graycomment{'2 possibilities: (1) Where an autistic person feels comfortable. Communication is plain, without irony or sarcasm. No idioms, no metaphors. Very precise. (2) Learning effect. Model does not adapt to autistic communication. Ambiguous communication. Angry face, but everything is fine. Contradictory signals. Ironic, sarcastic.' - E3}

As a friend, the agent should share similar hobbies, interests, and experiences to enhance rapport.
Conversely, as a training partner, the agent should excel in handling ambiguous communication, including the use of metaphors and irony. The recommendation is to allow players to create a character at the beginning of the game for a more immersive experience and personalization.

\subsection{Communication}

The minimum communication standard should include a chat-based system, with verbal communication preferred. Communication should aim for human-likeness, incorporating sensory capabilities for the agent such as eyes and ears to interpret gestures and facial expressions, and facilitating verbal interaction.

\graycomment{'Main goal: As human-like as possible. Verbal, non-verbal. With eye contact. With feet, hands. Minimal goal: Text-based, written with a focus on the verbal. If 'Siri' voice is better text-based and written.' - E4}

\subsection{Visualization}

Experts are undecided on whether the visualization should lean towards realism or stylizing. One expert noted the potential risk associated with the Uncanny Valley phenomenon \cite{6213238}.

\graycomment{'Rather stylized. Because it is difficult to achieve realism. Uncanny Valley effect. Therefore, to use simpler and more abstract forms.' - E2}

\graycomment{'Realistic enough to express emotion and other nonverbal cues. Similar to a human.' - E4}

\subsection{Feedback and adaptation}

The agent should provide feedback on the conversation, clarifying the meaning of its statements to the user, especially in the proposed training scenario. In addition, there is a suggestion to gradually increase the difficulty level in the training scenario.

\graycomment{'Different difficulty levels, varying during "playing". Not linear, but with variation.' - E3}

Agents can also help by assigning quests for real-life activities. Afterwards, there can be a discussion to determine if the given quests were completed successfully.

\graycomment{'Regular check-ins. The agent can assign tasks for real-life scenarios, such as engaging in a conversation about a specific topic. During the next interaction, the agent verifies whether the task has been completed.' - E1}

Feedback on important points is valuable for improving performance, and to enhance motivation, rewards are suggested. This could be easily integrated with the provided quest system.

\graycomment{'Snapshot of important parts of the conversation.' - E7}

\graycomment{'Positive reinforcement. Mistakes are allowed.' - E5}

\graycomment{'Rewards such as collecting points, more scenarios, and higher difficulty levels.' - E6}

Another crucial aspect is the presence of proactive agents capable of taking the lead in communication when necessary, providing positive feedback, and fostering an environment where it is acceptable to make mistakes.

\graycomment{'How one would have helped them. Ask proactive questions.
Not loud, rather calm and composed. Understanding. Non-judgmental. Be specific.' - E2}

\subsection{Evaluation}

This section outlines our findings regarding usability, user motivation, and evaluation of agent performance. One crucial aspect of evaluating user performance is assessing the transfer of knowledge between simulated and real-life scenarios.

\graycomment{'How do I feel in this situation? Do I experience a transfer as in other situations? Have I remembered myself in a real situation that I had experienced with the agent?' - E1}

Additionally, questionnaires can assess user motivation and usability. Potential suggestions include measuring the Bot Usability Scale (BUS) \cite{Borsci2022} or using the System Usability Scale \cite{Brooke1995} to assess the overall usability of the system.

\graycomment{'Evaluate performance with Social Responsiveness Scale (SRS). Has an autistic person noticeably changed in their interactions with others?' - E1}

From a therapeutic perspective, the user's performance is assessed using the Social Responsiveness Scale (SRS) \cite{constantino2021social}.

\subsection{Summary}

This section summarizes our findings and addresses the research question. 

\begin{itemize}
\item RQ1: What characteristics should empathetic agents have? \newline
\textit{There are two possible solutions including (1) a conversational partner or (2) a training partner for real life scenarios.}
\item RQ2: How to visualize empathetic agents? \newline
\textit{Two possible solutions including (1) as realistic as possible or (2) stylized.}
\item RQ3: How to interact with empathetic agents? \newline
\textit{Minimum communication includes a chat-based system, while more realistic communication incorporates both verbal and non-verbal aspects, including sensors such as microphones and cameras.}
\item RQ4: How to evaluate success of empathetic agents? \newline
\textit{The questionnaire encompasses measurements such as the Bot Usability Scale, System Usability Scale, Social Responsiveness Scale, and a rating scale.}
\end{itemize}

\section {Discussion}

Large Language Models, coupled with the visualization of conversation partners, offer an opportunity to train and enhance social skills for people with ASD. According to our experts, two scenarios are feasible: (1) an agent acting as a proxy for a human conversation partner and (2) as an agent serving as a training partner. Incorporating game elements such as quests to guide users on what to discuss with the agents, or having the agent assign real-life quests to the user, aims to provide long-term motivation and steep learning curves. The agent can autonomously and proactive assign quests to the user prompting specific conversations in real life. At a later stage, the agent can check with the user regarding the progress and outcome of the real-life quest, possibly using a rating scale. This scale can also serve to measure learning progress over time. Moreover, implementing dynamic difficulty levels for practicing everyday conversation scenarios enhances engagement and motivation for improvement. The visualization is under discussion, with experts divided between realism and stylization, indicating the need for further research.

\bibliography{01_refs}

\begin{thebibliography}{10}

\bibitem{Ali2020}
M.~R. Ali, S.~Z. Razavi, R.~Langevin, A.~A. Mamun, B.~Kane, R.~Rawassizadeh, L.~K. Schubert, and E.~Hoque.
\newblock A virtual conversational agent for teens with autism spectrum disorder.
\newblock {\em Proceedings of the 20th ACM International Conference on Intelligent Virtual Agents}, 2020.

\bibitem{Bernardini2021}
S.~Bernardini, K.~Porayska-Pomsta, and H.~Sampath.
\newblock Designing an intelligent virtual agent for social communication in autism.
\newblock {\em Proceedings of the AAAI Conference on Artificial Intelligence and Interactive Digital Entertainment}, 9:9--15, 2021.

\bibitem{Borsci2022}
Simone Borsci, Alessio Malizia, Martin Schmettow, Frank van~der Velde, Gunay Tariverdiyeva, Divyaa Balaji, and Alan Chamberlain.
\newblock The chatbot usability scale: the design and pilot of a usability scale for interaction with ai-based conversational agents.
\newblock {\em Personal and Ubiquitous Computing}, 26:1--25, 02 2022.

\bibitem{Brooke1995}
John Brooke.
\newblock Sus: A quick and dirty usability scale.
\newblock {\em Usability Eval. Ind.}, 189, 11 1995.

\bibitem{constantino2021social}
John~N Constantino.
\newblock Social responsiveness scale.
\newblock In {\em Encyclopedia of autism spectrum disorders}, pages 4457--4467. Springer, 2021.

\bibitem{AmericanPsychiatricAssociation}
John Cooper.
\newblock Diagnostic and statistical manual of mental disorders (4th edn, text revision) (dsm-iv-tr).
\newblock {\em British Journal of Psychiatry - BRIT J PSYCHIAT}, 179, 07 2001.

\bibitem{Hiniker2013}
Alexis Hiniker, Joy Daniels, and Heidi Williamson.
\newblock Go go games: Therapeutic video games for children with autism spectrum disorders.
\newblock pages 463--466, 06 2013.

\bibitem{Karami2023}
B.~Karami, R.~Koushki, F.~Arabgol, M.~Rahmani, and A.~Vahabie.
\newblock Effectiveness of virtual/augmented reality–based therapeutic interventions on individuals with autism spectrum disorder: a comprehensive meta-analysis.
\newblock {\em Frontiers in Psychiatry}, 12, 2021.

\bibitem{MesaGresa2018}
P.~Mesa-Gresa, R.~Oltra-Badenes, J.~Lozano-Quilis, and J.~Gil-Gómez.
\newblock Effectiveness of virtual reality for children and adolescents with autism spectrum disorder: an evidence-based systematic review.
\newblock {\em Sensors}, 18:2486, 2018.

\bibitem{Milne2010}
M.~Milne, M.~H. Luerssen, T.~Lewis, R.~Leibbrandt, and D.~Powers.
\newblock Development of a virtual agent based social tutor for children with autism spectrum disorders.
\newblock {\em The 2010 International Joint Conference on Neural Networks (IJCNN)}, 2010.

\bibitem{6213238}
Masahiro Mori, Karl~F. MacDorman, and Norri Kageki.
\newblock The uncanny valley [from the field].
\newblock {\em IEEE Robotics \& Automation Magazine}, 19(2):98--100, 2012.

\bibitem{Ng2018}
Yiu-kai Ng and Maria Pera.
\newblock Recommending social-interactive games for adults with autism spectrum disorders (asd).
\newblock pages 209--213, 09 2018.

\bibitem{park2023generative}
Joon~Sung Park, Joseph~C. O'Brien, Carrie~J. Cai, Meredith~Ringel Morris, Percy Liang, and Michael~S. Bernstein.
\newblock Generative agents: Interactive simulacra of human behavior, 2023.

\bibitem{Rodrigues2014}
Sérgio Rodrigues, Samuel Mascarenhas, João Dias, and Ana Paiva.
\newblock A process model of empathy for virtual agents.
\newblock {\em Interacting with Computers}, 27, 02 2014.

\bibitem{Zhang2024}
C.~Zhang, K.~Yang, S.~Hu, Z.~Wang, G.~Li, Y.~Sun, C.~Zhang, Z.~Zhang, A.~Liu, S.~Zhu, X.~Chang, J.~Zhang, F.~Yin, Y.~Liang, and Y.~Yang.
\newblock Proagent: building proactive cooperative agents with large language models.
\newblock {\em Proceedings of the AAAI Conference on Artificial Intelligence}, 38:17591--17599, 2024.

\end{thebibliography}

\end{document}